\documentclass[aps,twocolumn,groupedaddress,showpacs,prb,floatfix]{revtex4}
\usepackage{graphicx,rotating,subfigure,amsmath,amsfonts,amssymb,delarray}
\renewcommand{\vec}[1]{\boldsymbol #1}
\newcommand{\e}{\text{e}}
\newcommand{\im}{\text{i}}

\begin{document}
\bibliographystyle{apsrev}


\title{Thermodynamics of a one-dimensional $S=1/2$ spin-orbital model}


\author{J. Sirker}
\email[]{sirker@phys.unsw.edu.au}
\affiliation{School of Physics, The University of New South Wales,
  Sydney 2052, Australia}


\date{\today}

\begin{abstract}
The thermodynamic properties of a one-dimensional model describing
spin dynamics in the presence of a twofold orbital degeneracy are
studied numerically using the transfer-matrix renormalization group
(TMRG). The model contains an integrable $SU(4)$-symmetric point and a
gapless phase which is $SU(4)$ invariant up to a rescaling of the
velocities for spin and orbital degrees of freedom which allows
detailed comparison of the numerical results with conformal field
theory. We pay special attention to the correlation lengths which show
an intriguing evolution with temperature. We find that the model shows
an intrinsic tendency towards dimerization at finite temperature even
if the ground state is not dimerized. 
\end{abstract}
\pacs{75.10Jm, 05.70.-a, 05.10.Cc}

\maketitle

\section{Introduction}
\label{Intro}
Contrary to a usual band insulator the internal degrees of freedom
remain still active in a {\it Mott insulator}. In most of these
insulators the spatial distribution of electrons around every atom is
frozen in at the melting point and changes little down to zero
temperature so that the spin is the only active internal degree of
freedom. This leads to the widely studied spin models as for example
the Heisenberg model or the t-J model when hole-doping away from the
insulating case is studied. In recent years, however, there has been
considerable interest in systems which have additional {\it orbital
degrees of freedom} so that the spatial distribution of electrons
(orbital order) may change considerably with
temperature.\cite{KugelKhomskii,TokuraNagaosa} Transition-metal oxides
where orbitals are believed to play an important role are for example
LaMnO$_3$ \cite{TokuraCMR,TokuraNagaosa,CoeyViret}, LaTiO$_3$
\cite{KhaliullinMaekawa,KeimerCasa,HembergerKrug,CwikLorenz,MochizukiImada}
or YVO$_3$
\cite{KhaliullinHorsch,RenPalstra,RenPalstra2,NoguchiNakazawa,SirkerKhaliullin,KeimerSirker}.
In such systems a large number of states may be nearly degenerate
leading to very unusual magnetic properties as for example {\it
colossal magnetoresistance} \cite{TokuraCMR} or {\it magnetization
reversals} with temperature \cite{RenPalstra,RenPalstra2} but making
on the other hand a microscopic description extremely difficult.

\begin{figure}[ht]
\begin{center}
  \includegraphics*[width=0.9\columnwidth,height=0.25\textheight]{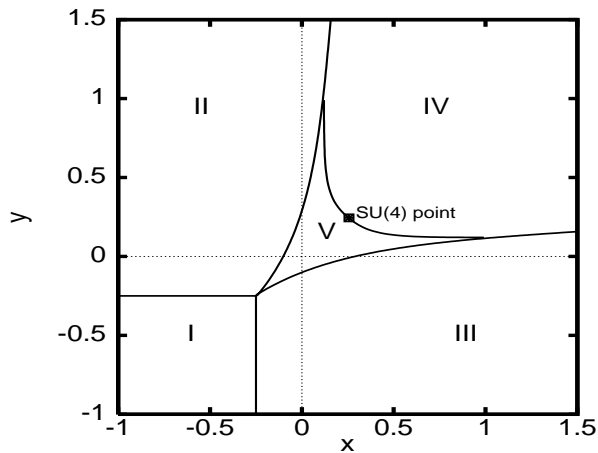}
\end{center}
\caption{Phase diagram for the model (\ref{eq2}) taken from
  Ref.~\onlinecite{ItoiQin}. Five phases exist as described in the
  main text.}
\label{Su4.fig0}
\end{figure}

From a theoretical point of view most interest has focussed on
one-dimensional (1D) versions of the {\it Kugel-Khomskii
Hamiltonian}\cite{KugelKhomskii} which allow for a full quantum
mechanical treatment. The simplest and most symmetric model describing
an $S=1/2$ spin system with a twofold orbital degeneracy is derived if
anisotropies in the orbital sector which are present due to Hund's
rule coupling are completely ignored. In this case the Hamiltonian is
given by
\begin{equation}
\label{eq1}
H = \sum_i \left(\vec{S}_i\vec{S}_{i+1} +
\frac{1}{4}\right)\left(\vec{\tau}_i\vec{\tau}_{i+1}+\frac{1}{4}\right)
\end{equation}
where $\tau=1/2$ is an {\it orbital pseudospin} describing the
occupation of two degenerate orbitals. The model has not only the
obvious $SU(2)\times SU(2)$ symmetry, i.e.~$SU(2)$ symmetry in both
spin and pseudospin space, but also $SU(4)$ symmetry unifying the spin
and orbital degrees of freedom.\cite{LiMa,ItoiQin} The model is
equivalent to the integrable $q=4$ Uimin-Sutherland model, which has
been solved by Bethe ansatz.\cite{Sutherland} The exact ground state
and the excitation spectrum consisting of three gapless mixed
spin-orbital modes has been obtained. Thermodynamic properties of the
$SU(4)$ symmetric model in 1D have been investigated by the quantum
Monte Carlo method.\cite{FrischmuthMila} In a real compound, however,
symmetry will always be reduced due to Hund's rule coupling.
A widely studied toy model to investigate the physical consequences of
symmetry breaking is given by
\begin{equation}
\label{eq2}
H=\sum_i (\vec{S}_i\vec{S}_{i+1} + x)(\vec{\tau}_i\vec{\tau}_{i+1} + y) 
\end{equation}
with real constants $x$,
 $y$.\cite{LiMa,YamashitaShibata,PatiSingh,FrischmuthMila,ItoiQin,AzariaBoulat,PatiSinghKhomskii}
 The model is $SU(2)\times SU(2)$ symmetric and exhibits an additional
 $Z_2$ symmetry, interchanging spin and orbital degrees of freedom, if
 $x=y$. The ground-state phase diagram depending on $x$, $y$ shows 5
 phases \cite{AzariaBoulat,PatiSingh,PatiSinghKhomskii,ItoiQin}: a
 phase with fully polarized ferromagnetic spins and orbitals (I),
 phases with antiferromagnetic spin and ferromagnetic orbital states
 (II) and vice versa (III), a dimerized phase (IV) and a gapless phase
 (V) which shows $SU(4)$ symmetry at low energies up to a rescaling of
 the spin and orbital velocities (see Fig.~\ref{Su4.fig0}).

In this paper we will study thermodynamic properties of the
Hamiltonian (\ref{eq2}) at the $SU(4)$-symmetric point, at $x=0.1$,
$y=0.25$ and at $x=0.5$, $y=0.5$ which are representative for the
gapless phase (V) and for the dimerized phase (IV), respectively. We
will focus on two aspects: First we will show that all thermodynamic
data at low temperatures (low-T) are consistently described by
conformal field theory (CFT) if the system is critical at
$T=0$. Second we will show that the model is highly susceptible for
dimerization at {\it finite temperature} even if the ground state is
not dimerized. The outline is as follows: In section \ref{TMRG} we
will briefly introduce the transfer-matrix renormalization group
(TMRG) which is the numerical method used to study the model at finite
temperatures. We will discuss our results for the $SU(4)$-symmetric
case in section \ref{Su4} before we turn to the point $x=0.1$,
$y=0.25$ (gapless phase) in \ref{gapless} and to $x=0.5$, $y=0.5$
(dimerized phase) in \ref{dimerized}. A short summary and some
conclusions are given in section \ref{Conclusion}.

\section{Transfer-matrix renormalization group}
\label{TMRG}
The idea of the TMRG is to express the partition function $Z$ of a
one-dimensional quantum model by that of an equivalent two-dimensional
classical model which can be derived by the Trotter-Suzuki
formula.\cite{Trotter,Suzuki2} For the classical model a suitable
transfer-matrix can be defined which allows for the calculation of all
thermodynamic quantities in the {\it thermodynamic limit} by
considering solely the largest eigenvalue of this transfer
matrix. Here we will use the Trotter-Suzuki mapping introduced in
Refs.~\onlinecite{SirkerKluemperEPL,SirkerKluemperPRB} which yields
\begin{eqnarray}
\label{TMRG.eq1}
Z &=& \lim_{M \rightarrow
  \infty}\text{Tr}\left\{\left[T_1(\epsilon)T_2(\epsilon)\right]^{M/2}\right\}
  \nonumber \\
  \mbox{with} && T_{1,2}(\epsilon) = T_{R,L}\exp\left[-\epsilon H + \mathcal{O}(\epsilon^2)\right]
\end{eqnarray}
where $\epsilon=\beta/M$,\footnote{Within this paper we will use
  $\epsilon=0.05$ in all calculations.} $\beta$ being the inverse
  temperature and $M$ an integer number. $T_{R,L}$ denotes the right-
  and left-shift operator, respectively. This mapping allows for the
  definition of a transfer matrix with repeat length 1 and the free
  energy in the thermodynamic limit is then given by
\begin{equation}
\label{TMRG.eq2}
f_{M} = -T \ln \Lambda_0 \; ,
\end{equation}
where $\Lambda_0$ is the largest eigenvalue of the column-to-column
transfer matrix $T_M$. This eigenvalue is non-degenerate and real for
any finite temperature. Correlation lengths (CLs) $\xi_n$ are also easy
to calculate within this approach. They are given by
\begin{equation}
\label{TMRG.eq3}
\xi_n^{-1} = \ln \left| \frac{\Lambda_0}{\Lambda_n} \right| \quad , \quad k_n = \arg \left( \frac{\Lambda_n}{\Lambda_0} \right) \; .
\end{equation} 
where $\Lambda_n$ are next-leading eigenvalues of $T_M$ and $k_n$ is
the corresponding wave vector. The appearance of in principle
infinitely many CLs can be understood as follows: Every two-point
correlation function can be expanded as 
\begin{equation}
\label{TMRG.eq4}
\langle O_1\,
O_r\rangle -\langle O_1\rangle\langle O_r\rangle = \sum_n M_n \,
\e^{-r/\xi_n}\,\e^{\im k_n r} 
\end{equation} 
with {\it matrix elements} $M_n$ where $O_i$ denotes some operator at
site $i$. The long distance behavior is dominated by the CL
$\xi_\alpha$ belonging to the largest eigenvalue $\Lambda_\alpha$
($\alpha\neq 0$) which satisfies the condition $M_\alpha \neq 0$. Note
that several CLs $\xi$ with the same wave vector $k$ can appear in this
asymptotic expansion. In the structure factor each exponential term
yields an (in principle) measurable Lorentz function with center at
$k_n$, height $\sim M_n \,\xi_n/\pi$ and width $\sim 2/\xi_n$. The
sharpest peak corresponds to the leading instability towards the onset
of long range order and hence, a crossover in the leading CL indicates
a change of the nature of the long range order. The transfer matrix is
enlarged in imaginary time direction (being equivalent to a decrease
in temperature) by using a density matrix renormalization group (DMRG)
algorithm. In all following calculations we have always retained
between 350 and 512 states. For details of the algorithm the reader is
referred to
Refs.~\onlinecite{Peschel,WangXiang,Shibata,SirkerKluemperEPL}.
\section{The $SU(4)$ symmetric point}
\label{Su4}
Although thermodynamic properties of the $SU(4)$ symmetric model have
already been studied using quantum Monte Carlo
algorithms,\cite{FrischmuthMila} we will start with this special point
for several reasons. First, it provides a good test for the numerics
because a comparison with Bethe ansatz and CFT results is directly
possible. Second, there are still interesting properties of this model
at finite temperature which have not been addressed so far: By
comparing the numerically calculated specific heat, spin
susceptibility and the correlation length data we will show that even
in this highly symmetric case the system is susceptible for
dimerization at finite temperature. We will also show that CFT
describes consistently all data in the low temperature limit.

\begin{figure}[ht]
\begin{center}
  \includegraphics*[width=0.98\columnwidth]{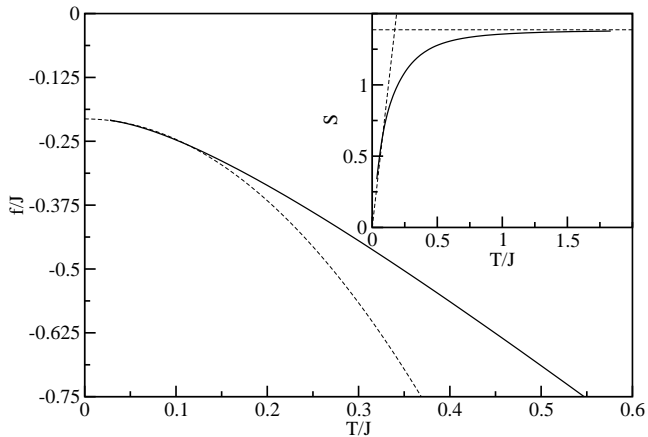}
\end{center}
\caption{Free energy (solid line) 
  and low-T asymptotics (dashed line) $f\sim e_0-4T^2$ as expected
  from CFT with $e_0$ as specified in the text. The inset shows the
  entropy (solid line) with dashed lines denoting the low ($S\sim 8T$)
  and high-T asymptotics ($S\sim \ln 4$).}
\label{Su4.fig1}
\end{figure}

\begin{figure}[ht]
\begin{center}
  \includegraphics*[width=0.98\columnwidth]{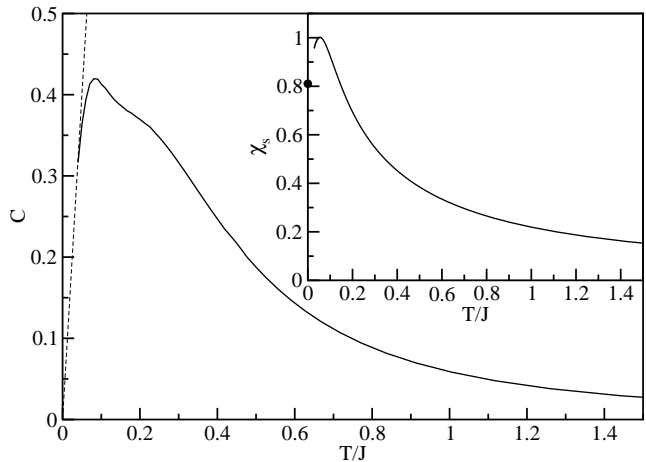}
\end{center}
\caption{Specific heat as a function of temperature (solid line) where
  the dashed line denotes again the low-T asymptotic behavior $C\sim
  8T$. In the inset the spin susceptibility is shown and the dot
  denotes the value expected from CFT.}
\label{Su4.fig2}
\end{figure}

In Fig.~\ref{Su4.fig1} the free energy $f$ calculated by the TMRG
method is shown. 
The low-energy properties of the model are known to be described by
the universality class of the $SU(4)_1$ Wess-Zumino-Witten (WZW)
model,\cite{Affleck_SU(N),ItoiQin} equivalent to 3 free boson modes so
that we expect from CFT in the low-T limit
\begin{equation}
\label{SU4.eq1}
f=e_0-\frac{\pi c}{6 v}T^2
\end{equation}
with {\it central charge} $c=3$. The velocity of the elementary
excitations $v=\pi/8$ and the ground state energy $e_0\approx
-0.20628$ are known from BA.\cite{Sutherland} Eq.~(\ref{SU4.eq1}) is
in excellent agreement with the numerics for $T/J<0.1$. The inset of
Fig.~\ref{Su4.fig1} displays the entropy $S=-\partial f/\partial T$
which should be given in the low-T limit by $S\sim 8T$
according to Eq.~(\ref{SU4.eq1}) and by $S\sim\ln 4$ at high
temperatures also in good agreement with the numerics.
The specific heat calculated by taking the numerical derivative
$C=-T\partial^2 f/\partial T^2$ is shown in Fig.~\ref{Su4.fig2}.
Somewhat surprisingly we find apart from a pronounced peak at
$T/J\approx 0.08$ an additional shoulder at $T/J\approx 0.22$. We will
return to this point soon but first discuss the spin susceptibility
$\chi_s$\footnote{Note that spin and orbital susceptibility are equal
along the line $x=y$.} at magnetic field $h=0$ which is shown in the
inset of Fig.~\ref{Su4.fig2}. This quantity is calculated numerically
by $\chi_s|_{h=0}= m/h$ where $h$ is a small applied magnetic field of
the order $J/100$ and $m$ is the corresponding magnetization which can
be calculated directly within TMRG.\cite{Peschel} To calculate
$\chi_s(T=0)$ within CFT one has to remind that the $SU(4)_1$ WZW
model is equivalent to a sum of 2 decoupled $SU(2)_2$ WZW
models.\cite{Affleck_SU(N),ItoiQin} This model is therefore a rare
example of a critical spin model with Kac-Moody central charge
$k>1$. It has been conjectured\cite{Affleck_WZW} that the
susceptibility of an isotropic spin-s antiferromagnet will be given by
$\chi_s=k/2\pi v_s$ so that we expect $\chi_s=1/\pi v_s=8/\pi^2$ which
is in good agreement with a linear extrapolation of the numerical
result to $T=0$.

By comparing $C$ and $\chi_s$ which show both a peak in the same
temperature region $T/J\approx 0.08$ we could identify this peak as
due to the elementary excitations 
of the $SU(4)_1$ WZW model. Note also that this peak marks the point
in temperature where the asymptotic expressions from CFT shown in
Figs.~\ref{Su4.fig1} and \ref{Su4.fig2} become invalid. To understand
the appearance of the shoulder we note that the energy for a single
bond is given by $-J/4$ if the spins form a singlet (triplet) and the
orbitals a triplet (singlet) and by $+J/4$ if both are in singlet or
triplet configuration. Therefore the excitation energy $\delta$ for a
single bond is $\delta=J/2$. In thermodynamic data, however, it makes
no difference for unbound excitations if they appear in pairs or not
so that the energy scale is set by $\delta/2=J/4$. This approximately
coincides with the shoulder in Fig.~\ref{Su4.fig2}. So we may
attribute this structure to the excitations of a chain with short
range dimer order in both spin and orbital sector which seems to form
at finite temperature although the ground state is not dimerized. This
picture is further supported by an investigation of the leading CLs
which are plotted as $T\xi/J$ in Fig.~\ref{Su4.fig3} because a $1/T$
divergence at low-T of those CLs belonging to the critical modes is
expected (see below).
\begin{figure}[ht]
\begin{center}
  \includegraphics*[width=0.98\columnwidth]{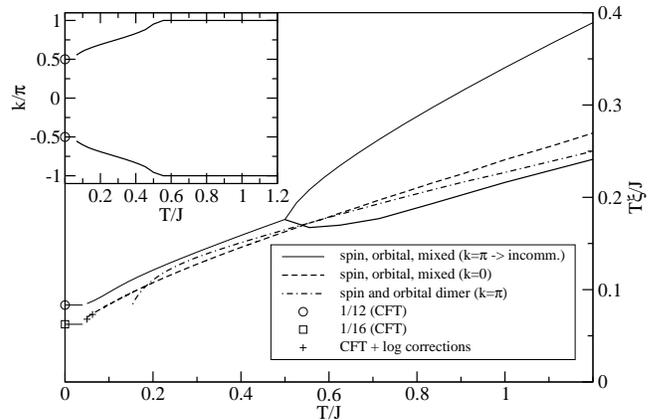}
\end{center}
\caption{Leading correlation lengths at the $SU(4)$-point. 
The symbols at $T=0$ are CFT results where the short lines should
indicate that these results are expected to be valid (up to
log.~corrections) also at small but finite temperatures. The inset
shows the temperature dependence of the wave vector $k$ in the case of
incommensurate oscillations.}
\label{Su4.fig3}
\end{figure}
Here a CL $\xi_D$ is present (dot-dashed line) which appears in the
asymptotic expansion (Eq.~(\ref{TMRG.eq4})) of the dimer correlation
functions
$\langle S^z_0S^z_1 S^z_rS^z_{r+1}\rangle$ and 
$\langle \tau^z_0\tau^z_1 \tau^z_r\tau^z_{r+1}\rangle$. $\xi_D$ is in
$T/J\in [0.2:0.5]$ only slightly smaller than the largest CL
supporting the picture of a short range dimer order in this
temperature regime. For lower temperatures $T\xi_D$ starts to decrease
as is expected because the ground state is not dimerized and $\xi_D$
should therefore remain finite.
Note that due to the $SU(4)$ symmetry the relation
\begin{equation}
\label{SU4.eq2}
\langle S_i^\alpha S_j^\alpha \rangle = \langle \tau_i^\alpha
\tau_j^\alpha \rangle = 4\langle S_i^\alpha \tau_i^\beta S_j^\alpha
\tau_j^\beta \rangle
\end{equation}
is valid for the thermodynamic correlation functions with arbitrary
components $\alpha,\beta = x,y,z$.\cite{LiMa} Therefore the CLs
depicted in Fig.~\ref{Su4.fig3} as straight and dashed lines have
connected non-zero matrix elements for the spin-spin, orbital-orbital
and mixed spin-orbital correlator. At high temperatures these CLs show
commensurate oscillations, i.e., the wave vector $k$ is equal $0$ or
$\pi$. The 2 CLs which show $\pi$-oscillations at high temperatures
(straight lines in Fig.~\ref{Su4.fig3}) merge at a temperature
$T/J\sim 0.5$ with a square root singularity (note that the number of
data points is relatively small in this region) and the oscillations
become incommensurate and temperature dependent at lower temperatures
as shown in the inset of Fig.~\ref{Su4.fig3}. A similar behavior has
been observed for the longitudinal spin CL in the $XXZ$-model for
finite magnetic field.\cite{KluemperScheeren} Note that incommensurate
oscillations are associated with a pair of complex conjugate
eigenvalues of the transfer matrix (see Eq.~(\ref{TMRG.eq3})) which
can occur because the transfer matrix is real but
non-symmetric. Additionally a CL (dashed line) is present which is
non-oscillating for all temperatures.

It is known that the energy levels $E_i$ of a finite chain of length
$L$ show a finite-size scaling behavior
\begin{equation}
\label{SU4.eq3}
E_i-E_0 = \frac{2\pi v x_i}{L}
\end{equation}
where $E_0$ is the ground state energy and $x_i$ is the scaling
dimension of the operator leading to the excited state with energy
$E_i$.\cite{Cardy84} We can get the corresponding result for an
infinite chain at {\it finite} temperature by the substitution
$L\rightarrow v/T$. This leads to
\begin{equation}
\label{SU4.eq4}
\xi_i^{-1} = \ln\Lambda_0-\ln\left|\Lambda_i\right| = \frac{2\pi x_i T}{v}
\end{equation} 
where $\Lambda_i$, $\Lambda_0$ are eigenvalues of the transfer
matrix. The possible scaling dimensions for an $SU(N)$ symmetric model
are given by $x_p = p(N-p)/N$ with $p=1,2,\cdots,N-1$. The dominant
mode belongs to $p=1$ and oscillates at zero temperature with
$k=2\pi/N$.\cite{Affleck_SU(N),MajumdarMukherjee} In the case of
$SU(4)$, scaling dimensions $3/4,1,3/4$ are therefore possible and the
dominant mode has $x=3/4$ and oscillates with $k=\pi/2$. Using
Eq.~(\ref{SU4.eq4}) we obtain $\xi_{\pi/2}=1/12T$. The non-oscillating
CL must belong to the second possible scaling dimension $x=1$ leading
to $\xi_0 = 1/16T$ in the low-T regime. However, this $1/T$ behavior
will be directly observable only at very low temperatures (see
Ref.~\onlinecite{FrischmuthMila}). At the lowest temperatures we are
able to access in this study it is covered by logarithmic corrections
which are present here due to marginally irrelevant operators in the
low-energy effective theory.\cite{ItoiQin,MajumdarMukherjee}
Cardy has shown that such logarithmic corrections have universal
character and can be displayed in leading order as
\begin{equation}
\label{SU4.eq5}
E_i - E_0 = \frac{2\pi v}{L}\left(x_i +\frac{2b_n}{b}\frac{1}{\ln L}\right)
\end{equation} 
where $b_n$, $b$ are universal numbers.\cite{Cardy86} However, it
should be stressed that this asymptotic form is only observable for
very large chain lengths $L$ when the logarithmic correction becomes
independent of the coupling constant $g$ of the considered marginal
operator. Otherwise we can only replace $g$ by its renormalization
group improved value $g(\ln L)$ so that we must substitute $\ln
L\rightarrow \ln(L/L_0)$ in Eq.~(\ref{SU4.eq5}) with a non-universal
$L_0$. But even in this case Eq.~(\ref{SU4.eq5}) remains useful
because it describes the scaling behavior of {\it different} excited
states by a single parameter. By $L\rightarrow v/T$ we find for the
CLs
\begin{equation}
\label{SU4.eq6}
\xi_i^{-1} = \frac{2\pi T}{v}\left(x_i +\frac{2b_n}{b}\frac{1}{\ln T_0/T}\right)
\end{equation} 
where $T_0$ is non-universal. This means that logarithmic corrections
lead to an effective temperature dependent correction of the scaling
dimension $x_i$. The universal factor $2b_n/b$ for the $SU(4)$ model
has been derived in Ref.~\onlinecite{MajumdarMukherjee} and is given
by $2b_n/b=-1/16$ for scaling dimension $x=3/4$ and by $2b_n/b=-1/4$
for $x=1$. Therefore the logarithmic correction for $\xi_0$ is more
important than that for $\xi_{\pi/2}$ in qualitative agreement with
the data. To illustrate more quantitatively that the logarithmic
corrections can explain the deviations from the expected CFT values,
we have estimated $T_0$ at $T=0.05J$ and $T=0.0625J$ for $\xi_{\pi/2}$
by applying Eq.~(\ref{SU4.eq6}) and used this value to calculate
$\xi_0$ at the same temperature. The results are depicted as
plus-symbols in Fig.~\ref{Su4.fig3} and are in good agreement with the
numerically calculated $\xi_0$. We like to remark that this is a very
sensitive test because $T_0$ depends exponentially on $\xi$. Note also
that we cannot extend this comparison between field theory and
numerics to larger temperatures because we already know that the low
energy effective theory breaks down at least at $T/J\approx 0.08$.

\section{Gapless phase ($SU(4)$ with rescaled velocities)}
\label{gapless}
By an RG analysis it has been shown that the Hamiltonian (\ref{eq2})
is not only critical at the $SU(4)$-symmetric point but rather
exhibits an extended region in phase space around this point which is
gapless (phase V in Fig.\ref{Su4.fig0}).\cite{ItoiQin,AzariaBoulat}
Considering the parameters $x=1/4+\delta x$, $y=1/4+\delta y$ in
Eq.~(\ref{eq2}), where $\delta x$, $\delta y$ are small perturbations,
the findings can be summarized as follows: (1) For $\delta x = \delta
y <0$ the breaking of the $SU(4)$ symmetry down to $SU(2)\times SU(2)$
is irrelevant and the low-energy behavior is still described by the
$SU(4)_1$ WZW model. (2) $\delta x + \delta y < 0$ with $\delta x \neq
\delta y$ is marginally irrelevant and the critical theory is the
$SU(2)_2\times SU(2)_2$ WZW model with {\it unequal} velocities $v_s$,
$v_o$, but with unchanged scaling dimensions. (3) $\delta x + \delta y
> 0$ is marginally relevant leading to a Kosterlitz-Thouless type
transition into a dimerized phase (phase IV in Fig.\ref{Su4.fig0}).

\begin{figure}[ht]
\begin{center}
  \includegraphics*[width=0.98\columnwidth]{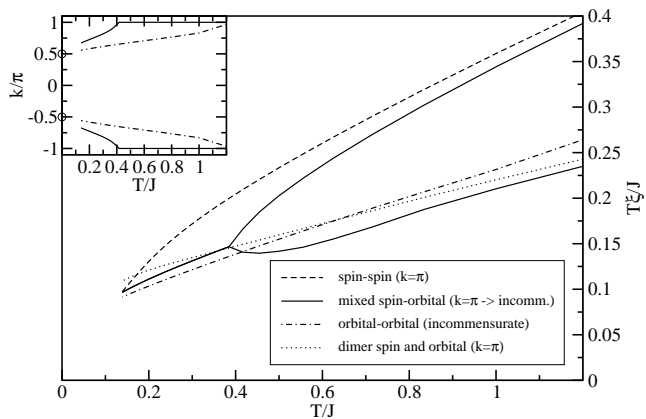}
\end{center}
\caption{Some of the leading spin and orbital correlation lengths for
  $(x,y)=(0.1,0.25)$. The wave vectors are shown in the inset if
  incommensurate oscillations occur. The circles in the inset denote
  $\pm \pi/2$.}
\label{gapless.fig1}
\end{figure}

\begin{figure}[ht]
\begin{center}
  \includegraphics*[width=0.98\columnwidth]{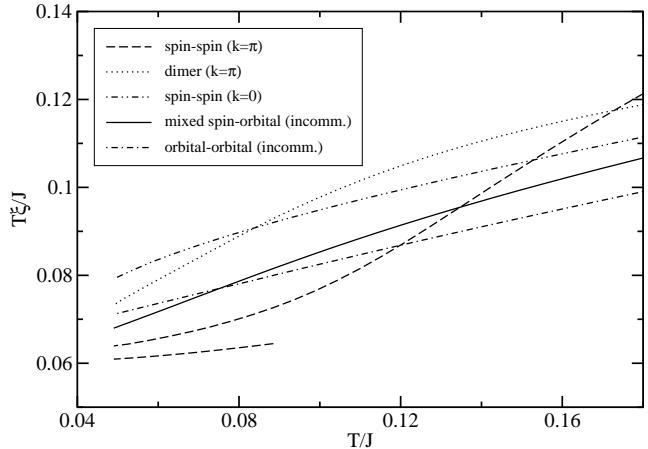}
\end{center}
\caption{Leading spin and orbital correlation lengths for
  $(x,y)=(0.1,0.25)$ at low temperatures.}
\label{gapless.fig1.1}
\end{figure}

Representatively for the gapless phase with unequal velocities of the
elementary excitations we will study the point $x=0.1$, $y=0.25$. Due
to the different velocities $v_s\neq v_o$ relation (\ref{SU4.eq2}) is
no longer valid and the degeneracy between spin-spin, orbital-orbital
and mixed spin-orbital CLs is lifted. This makes the situation far
more complex than in the $SU(4)$ case with numerous crossovers between
different CLs occurring in the non-universal regime at elevated
temperatures. To keep it concise we show only a selection of CLs at
high temperatures in Fig.~\ref{gapless.fig1} and show the 6 largest
CLs at the lowest numerically accessible temperatures separately in
Fig.~\ref{gapless.fig1.1}.
Over a wide temperature range a $\pi$-oscillating spin CL is
leading. This is expected because antiferromagnetic spin correlations
are stabilized with decreasing $x$. As in the $SU(4)$-symmetric case
two CLs with $\pi$-oscillations merge at a temperature $T/J\sim 0.4$
and show incommensurate oscillations below which approach $\pi/2$ at
$T=0$. However, in the $SU(4)$ case these CLs are threefold degenerate
belonging to spin, orbital and mixed spin-orbital correlation
functions. Here no degeneracy is present and the CLs only belong to
the mixed spin-orbital correlator. The orbital CL which shows $\pi/2$
oscillations at $T=0$ develops incommensurability already at higher
temperatures as shown in the inset of
Fig.~\ref{gapless.fig1}. Hitherto we have ignored the dimer CL $\xi_D$
which is not only one of the largest CL as in the $SU(4)$ case but is
indeed leading in a certain temperature range. This is a little bit
astonishing because we have moved away from the dimer phase by $\delta
x <0$. It shows that the spin-orbital model has an intrinsic tendency
towards dimerization at finite temperature even in regions of phase
space where the ground state is not dimerized. For a system consisting
of weakly coupled spin-orbital chains this suggests that long range
dimer order can be stabilized at finite temperature and that such a
system may show a phase transition from a dimer phase in this
temperature regime to a non-dimerized phase at lower temperatures. In
the limit $T\rightarrow 0$ we expect that $\xi_D$ remains finite and
Fig.~\ref{gapless.fig1.1} shows indeed a decreasing $T\xi_D$. However,
even at the lowest numerically accessible temperature $\xi_D$ is still
the second largest CL.
Dominating at low temperatures is a non-oscillating spin
CL. Furthermore we see a crossover between the mixed spin-orbital and
the orbital-orbital CL with incommensurate oscillations with the
orbital-orbital CL becoming the larger one at low-T. From CFT we also
expect a spin-spin CL with incommensurate oscillations which seems to
be missing in our data. We expect that the 2 spin-spin CLs with
$k=\pi$ depicted as dashed lines in Fig.~\ref{gapless.fig1.1} will
merge at lower temperature yielding this missing CL. This implies that
we still have not reached the temperature regime where the behavior of
the CLs becomes universal and can be described by CFT. Therefore we
cannot compare the numerics with CFT in detail. However, assuming that
we are close to the conformal regime and ignoring logarithmic
corrections completely we can extrapolate all CLs (except of the dimer
CL) shown in Fig.~\ref{gapless.fig1.1} starting from the numerical
data at the lowest accessible temperature according to their expected
$1/T$ behavior. This allows us to show that the data are in rough
agreement with CFT.

The non-oscillating spin-spin CL must belong to an excitation with
$x_s=1$ so that $\xi = v_s/2\pi T$ leading to $v_s\approx 0.5$. For
the corresponding non-oscillating orbital-orbital CL we find
numerically (data not shown) $\xi(T=0.047)\approx 0.86$ leading to
$v_o\approx 0.25$. 
We expect that the $SU(4)$ representation is decomposed symmetrically
into a spin and orbital sector both with central charge $c=3/2$ and
fundamental fields with scaling dimensions $x=3/8$. The only
difference between the two sectors are the velocities. All CLs with
$k=\pi/2$ at $T=0$ are therefore given by an excitation with $x_s =
x_o = 3/8$ so that we expect from CFT
\begin{equation}
\label{gapless.eq1}
\xi_{\pi/2}=\frac{1}{2\pi \frac{3}{8}(\frac{1}{v_s}+\frac{1}{v_o}) T} \sim \frac{0.071}{T}
\end{equation}
where we have substituted the velocities $v_s$, $v_o$ which we have
determined from the non-oscillating spin and orbital CL. At finite
temperatures the mixed, spin and orbital CL with incommensurate
oscillations differ from each other because each has different
logarithmic corrections.\cite{ItoiQin} 
We notice that the value estimated in Eq.~(\ref{gapless.eq1}) is
indeed in rough agreement with the numerical data at the lowest
accessible temperature (see Fig.~\ref{gapless.fig1.1}). The mixed
spin-orbital CL with $k=0$ is given by an excitation with $x_s = x_o =
1/2$ leading to
\begin{equation}
\label{gapless.eq2}
\xi_{0}^{s-o}=\frac{1}{2\pi \frac{1}{2}(\frac{1}{v_s}+\frac{1}{v_o}) T} \sim \frac{0.053}{T}
\end{equation}
predicting $\xi_{0}^{s-o}(T=0.047)\approx 1.12$ which is in good
agreement with the numerically calculated value
$\xi_{0}^{s-o}(T=0.047)\approx 1.09$ (data not shown).

\begin{figure}[ht]
\begin{center}
  \includegraphics*[width=0.98\columnwidth]{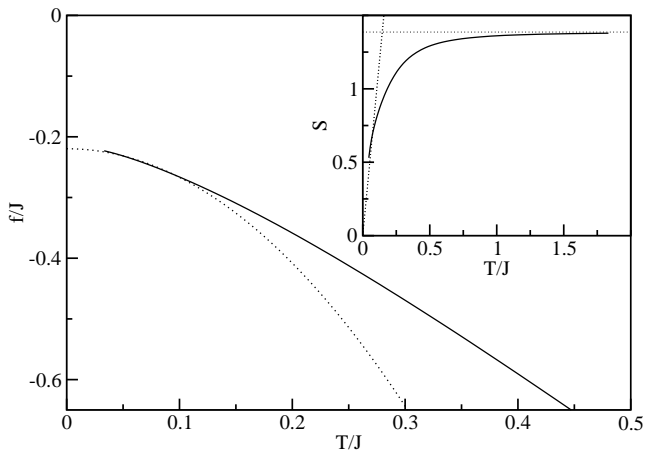}
\end{center}
\caption{Free energy and entropy (inset) at $(x,y)=(0.1,0.25)$ with
  low and high-T asymptotics (dotted lines) as described in the text.}
\label{gapless.fig2}
\end{figure}

\begin{figure}[ht]
\begin{center}
  \includegraphics*[width=0.98\columnwidth]{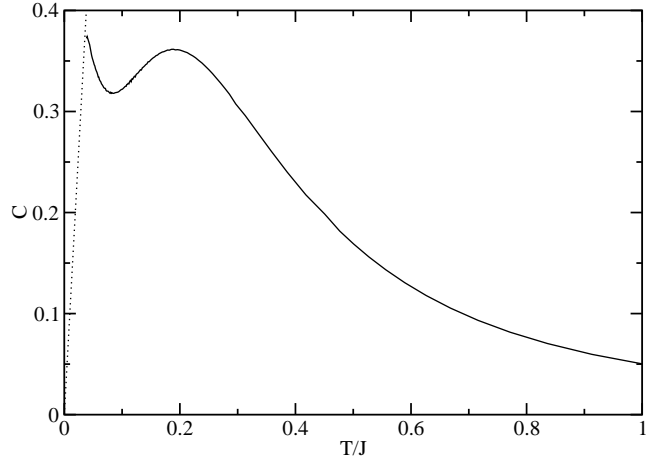}
\end{center}
\caption{Specific heat (solid line) and low-T asymptotics (dashed
  line) $C\sim3\pi T$ at $(x,y)=(0.1,0.25)$.}
\label{gapless.fig3}
\end{figure}

\begin{figure}[ht]
\begin{center}
  \includegraphics*[width=0.98\columnwidth]{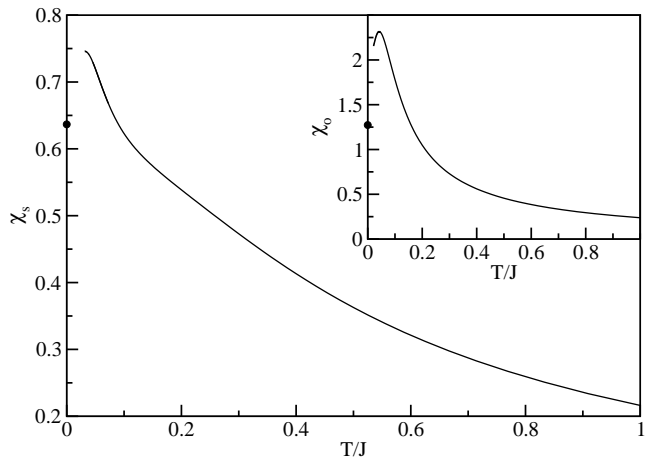}
\end{center}
\caption{Spin (main figure) and orbital (inset) susceptibilities at
$(x,y)=(0.1,0.25)$. The dots denote CFT results as described in the
text.}
\label{gapless.fig4}
\end{figure}

The CFT result for the free energy of the $SU(4)$-symmetric model
(Eq.~(\ref{SU4.eq1})) can be easily generalized to the case of
$SU(2)\times SU(2)$-symmetry with different velocities
\begin{equation}
\label{gapless.eq3}
f = e_0 -\frac{\pi c}{6}\left(\frac{1}{v_s}+\frac{1}{v_o}\right)T^2 
\end{equation}
where now $c=3/2$.
In Fig.~\ref{gapless.fig2} the numerically calculated free energy and
the low-T asymptotics according to this formula are shown. Here we
have used the spin and orbital velocities which we have
evaluated from the CLs and $e_0$ as a fitting parameter. 
We find $e_0\approx -0.2197 J$ and a good agreement with the numerical
data for $T/J<0.1$. In the inset of Fig.~\ref{gapless.fig2} the
entropy and the low-T ($S\sim 3\pi T$) and high-T asymptotics ($S\sim
\ln 4$) are shown. Here the low-T asymptotics is determined solely by
the spin and orbital velocities without any additional fitting
parameter showing that CFT for the $SU(2)\times SU(2)$ model with our
velocity estimates gives a consistent description of all thermodynamic
quantities at low energies. Next we want to consider the specific heat
$C$ shown in Fig.~\ref{gapless.fig3}.
Contrary to the $SU(4)$-symmetric case not only a peak and a shoulder
but instead two clearly distinguishable peaks are visible. Note that
due to the different velocities, orbital and spin excitations have now
different energy scales. However, any saturation of one type of
elementary excitations (spinons or orbitons) will inevitably destroy
also a description in terms of these elementary excitations for the
other sector due to the strong coupling between both. That this is
indeed the case is obvious from the susceptibility data shown in
Fig.~\ref{gapless.fig4} which show that spin and orbital
susceptibility are peaked approximately at the same
temperature.\footnote{More precisely, $\chi_o$ is peaked at $T\approx
0.043J$ and $\chi_s$ at $T\approx 0.033J$.}
 The second peak of $C$ is in a temperature range where short range
dimer order exists and may be attributed to excitations connected to
this. Finally we want to compare the susceptibility data with CFT. As
in the $SU(4)$ case we expect $\chi_{s,o}=k/2\pi v_{s,o}$ with
topological coupling constant $k=2$, however, now $v_s\neq v_o$. Using
the velocity estimates from the fit of the CLs we find $\chi_s\approx
2/\pi$ and $\chi_o\approx 4/\pi$ predicting $\chi_o$ at $T=0$ being
twice as large as $\chi_s$. Although we cannot access numerically
temperatures low enough to show that the susceptibilities converge to
these values as $T\rightarrow 0$ we note that $\chi_o$ is indeed more
than twice as large as $\chi_s$ at low temperatures.
\section{Dimerized phase}
\label{dimerized}
\begin{figure}[!ht]
\begin{center}
  \includegraphics*[width=0.98\columnwidth]{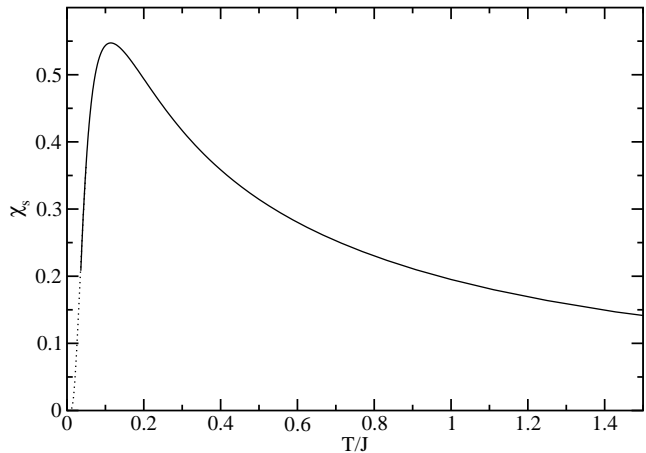}
\end{center}
\caption{Susceptibility (solid line) for $(x,y)=(0.5,0.5)$ where the
  dotted line denotes the best low-T fit with
  $S=0.491\exp(-0.09/T)/\sqrt T$.}
\label{dimer.fig1}
\end{figure}

As described in the previous section perturbing the $SU(4)$-symmetric
Hamiltonian by $\delta x + \delta y > 0$ is marginally relevant. As a
result a gap $\Delta$ in the excitation spectrum is expected. Here we
do not want to investigate this phase transition in detail but rather
restrict ourselves to study the point $x=0.5$, $y=0.5$
as a representative for the dimerized phase. First, consider the spin
susceptibility $\chi_s$ shown in Fig.~\ref{dimer.fig1}.
A sharp peak at $T=0.12J$ is visible. 
At lower temperatures $\chi_s$ goes to zero as expected for a gapped
system. The usual quadratic dispersion of elementary excitations above
the gap yields in 1D an expected low-T asymptotic behavior $\chi_s\sim
\exp(-\Delta/T)/\sqrt T$. Using this formula for a fit of the data in
the low-T regime we find $\Delta = 0.090\pm 0.005$ where we get a
certain error estimate by varying the fit interval.\footnote{This
error does not include systematic errors due to the truncation of the
Hilbert space which are difficult to estimate.} This value is in
agreement with a recent series expansion study.\cite{ZhengOitmaa}
\begin{figure}[!ht]
\begin{center}
  \includegraphics*[width=0.98\columnwidth]{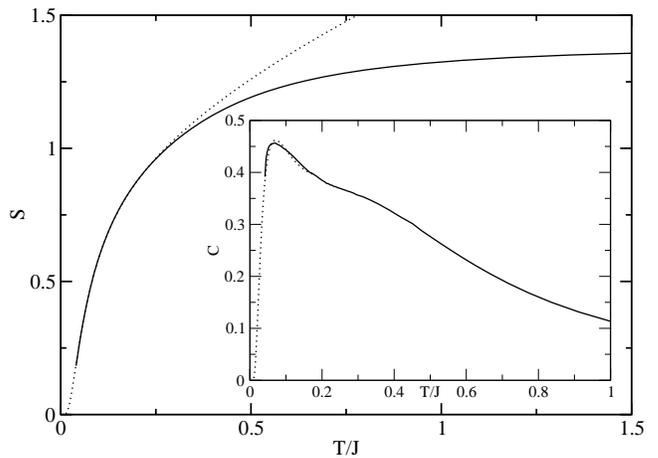}
\end{center}
\caption{Entropy and specific heat (inset) at $(x,y)=(0.5,0.5)$. The
  dashed lines denote low-T fits as described in the text.}
\label{dimer.fig2}
\end{figure}
In Fig.~\ref{dimer.fig2} the corresponding entropy $S$ is shown. For
the free energy we expect at low temperatures $f\sim
-T^{3/2}\exp(-\Delta/T)$ leading to $S\sim (A\sqrt T +B/\sqrt
T)\exp(-\Delta/T)$. Using $A, B, \Delta$ as parameter in a fit of the
numerical data for $T<0.1J$ we get the result shown as dotted line in
Fig.~\ref{dimer.fig2} which is an excellent fit even up to $T\sim
0.25J$. From this fit we find $\Delta=0.094\pm 0.004$ where the error
stems again from a variation of the interval used in the fit. We note
that this value is in perfect agreement with the estimate from the
susceptibility fit. In the inset of Fig.~\ref{dimer.fig2} the specific
heat (solid line) is shown. The dotted line denotes the low
temperature specific heat calculated from the entropy fit and shows
good agreement with the numerical data. Note that the excitation
energy $\delta$ for a single bond is now $J/4$ so that the peak in the
specific heat at approximately $\delta/2$ can be attributed again to
excitations of the chain with dimer order which will now become a true
long range order at $T=0$. Surprisingly we still find a shoulder in
$C$ at higher temperatures as in the $SU(4)$ case. Comparing with the
CLs plotted in Fig.~\ref{dimer.fig3} we find indeed that the
eigenspectrum of the transfer matrix at elevated temperatures has not
changed drastically compared to Fig.~\ref{Su4.fig3}.
\begin{figure}[!ht]
\begin{center}
  \includegraphics*[width=0.98\columnwidth]{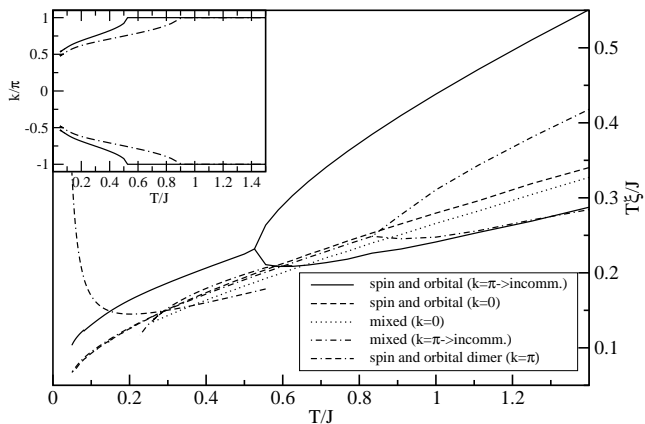}
\end{center}
\caption{Leading CLs at $(x,y)=(0.5,0.5)$. The inset shows the
  dependence of the wave vector on temperature in the case of
  incommensurate oscillations.}
\label{dimer.fig3}
\end{figure}
Especially, we find again a dimer CL (dot-double dashed line) which
becomes the second largest CL in $T/J\in [0.3:0.55]$ which coincides
with the shoulder in $C$. Whereas this dimer CL stays finite as
$T\rightarrow 0$ another dimer CL (also depicted as dot-dashed
line)\footnote{We use the same symbol for both CLs because they are of
the same type. However, they are connected to different eigenvalues
which are crossing at a certain temperature.} is diverging more
strongly than $1/T$ indicating long range dimer order at
$T=0$. Because $x=y$ the CLs belonging to the spin and orbital
correlation function are equal, however, CLs belonging to the mixed
correlator are different in general.\footnote{Eq.~(\ref{SU4.eq2}) is
not valid here.}  The two CLs (solid lines) which merge at $T/J\sim
0.5$ yielding a CL with corresponding incommensurate oscillations
below that temperature have indeed only non-zero matrix elements for
$\langle S^z_0 S^z_r\rangle$ and $\langle \tau^z_0
\tau^z_r\rangle$. The $\pi$-oscillating CLs with non-zero matrix
element for $\langle S^z_0\tau^z_0 S^z_r\tau^z_r\rangle$ merge already
at a higher temperature and the oscillations become incommensurate
below as shown in the inset of Fig.~\ref{Su4.fig3}. Similarly, the
largest non-oscillating CL belongs only to the spin and orbital
correlator whereas the second largest appears only in the asymptotic
expansion of the mixed correlator. The numerical data indicate that
all CLs apart from the dimer CL stay finite as $T\rightarrow 0$ which
is expected for a system with a gap.
\section{Conclusions}
\label{Conclusion}
In summary, we have used the TMRG method to study thermodynamic
properties of a spin-$1/2$ $SU(2)\times SU(2)$ spin-orbital model. We
have concentrated on three representative points in the phase diagram:
(1) The $SU(4)$ symmetric point where the model is Bethe-ansatz
integrable and shows critical properties at $T=0$, (2) a point where
the model still exhibits critical behavior at zero temperature,
however, the symmetry is reduced to $SU(2)\times SU(2)$, and (3) a
point where the ground state has long range dimer order.

In the first two cases we have compared our numerical results in
detail with CFT in the low temperature regime and with special
attention to the evolution of various CLs with temperature. In both
cases three different regimes could be identified: At elevated
temperatures crossovers between different CLs are frequent and the
corresponding wave vectors are commensurate. The $\pi$-oscillating CLs
belonging to the spin, orbital or mixed spin-orbital correlator merge
at a certain temperature with a square root singularity and the
oscillations become incommensurate and temperature dependent below
which characterizes the intermediate temperature regime. Finally, at
low temperatures the conformal invariant regime is reached where the
CLs belonging to critical excitations diverge as $1/T$ up to
logarithmic corrections and no crossover occur. The comparison of the
numerics with CFT was especially simple at the $SU(4)$ point because
the velocities of the elementary excitations are known exactly from
Bethe ansatz.\cite{Sutherland} In this case we have also been able to
show that the predicted logarithmic
corrections\cite{ItoiQin,MajumdarMukherjee} account quantitatively for
the observed deviations from a pure $\xi\sim 1/T$ behavior at low
temperatures. In the second case a comparison with CFT was far more
complicated as the velocities $v_s$, $v_o$ are a priori unknown and
several crossovers in the lowest numerically accessible temperature
region show that the conformal invariant regime was not
reached. However, we have been able to get rough estimates for the
spin and orbital velocities from the CLs and have shown that these
estimates give a consistent description of other thermodynamic
quantities within CFT.

For the point in the dimerized phase we have calculated the gap from
the susceptibility and entropy data and have found good agreement with
a recent series expansion study.\cite{ZhengOitmaa} More important, we
have found that the behavior of CLs in the high and intermediate
temperature regime here is quite similar to the gapless phase. At low
temperatures a dimer CL diverges more strongly than $1/T$ whereas all
other CLs stay finite as is expected for a system with long range
dimer order at $T=0$.

Most surprisingly we find in all 3 cases a strong tendency towards
dimerization in a certain finite temperature interval and dimerization
could be the leading instability in this region even if the ground
state is not dimerized. This behavior is very similar to what was
observed recently in another spin-orbital model with spin
$S=1$\cite{SirkerKhaliullin} indicating that such instability may be
an intrinsic property of these kind of Kugel-Khomskii Hamiltonians. In
a system of weakly coupled spin-orbital chains this instability may
turn into true long range dimer order at finite temperature. More
generally speaking, it indicates that even a pure electronic model may
show various temperature driven phase transitions between different
spin and orbital orderings due to the large number of nearly
degenerated states accessible to systems with unquenched orbital
degrees of freedom.
\begin{acknowledgments}
I thank A.~Kl\"umper and J.~Oitmaa for useful discussions. This work
has been partially supported by the DFG in SP1073 and by the
Australian Research Council. The numerical calculations were partially
performed using the APAC computing facilities.
\end{acknowledgments}

\end{document}